\documentclass[fleqn,10pt]{wlscirep}
\usepackage[utf8]{inputenc}
\usepackage[T1]{fontenc}

\usepackage[most]{tcolorbox}

\usepackage{amsmath}
\graphicspath{{Figures/}}

\title{Early Warning Signals for Bifurcations Embedded in High Dimensions}

\author[1,2,*]{Daniel Dylewsky}
\author[2]{Madhur Anand}
\author[1]{Chris T. Bauch}
\affil[1]{Department of Applied Mathematics, University of Waterloo, Waterloo, ON, Canada N2L 3G1}
\affil[2]{School of Environmental Sciences, University of Guelph, Guelph, ON, Canada N1G 2W1}

\affil[*]{ddylewsk@uwaterloo.ca}

\begin{abstract}
Recent work has highlighted the utility of methods for early warning signal detection in dynamic systems approaching critical tipping thresholds. Often these tipping points resemble local bifurcations, whose low dimensional dynamics can play out on a manifold embedded in a much higher dimensional state space. In many cases of practical relevance, the form of this embedding is poorly understood or entirely unknown. This paper explores how measurement of the critical phenomena that generically precede such bifurcations can be used to make inferences about some properties of their embeddings, and, conversely, how prior knowledge about the mechanism of bifurcation can robustify predictions of an oncoming tipping event. These modes of analysis are first demonstrated on a simple fluid flow system undergoing a Hopf bifurcation. The same approach is then applied to data associated with the West African monsoon shift, with results corroborated by existing models of the same system. This example highlights the effectiveness of the methodology even when applied to complex climate data, and demonstrates how a well-resolved spatial structure associated with the onset of atmospheric instability can be inferred purely from time series measurements.
\end{abstract}
\begin{document}

\flushbottom
\maketitle

\thispagestyle{empty}

\section*{Introduction}

Many dynamical systems contain critical thresholds which act as ``tipping points," parameter boundaries whose traversal induces an abrupt and potentially irreversible shift in the state. These critical transition events pose considerable concern in climatic, ecological, economic, and other domains in which a sudden, dramatic reconfiguration could have grave consequences for human welfare. Increasing attention has recently been devoted to understanding how early warning signals (EWS) associated with an oncoming tipping point may manifest in a system's dynamics before it reaches criticality.

Many such transitions are well described by bifurcation theory, which provides a mathematical foundation for how a smoothly-varied forcing on a system can elicit a potentially discontinuous change to its stable equilibria. However, while the most commonly occurring types of bifurcation play out in one or two dimensions, the systems in which we want to identify EWS often have many more degrees of freedom. In climate science, for example, where tipping phenomena are a subject of great interest, the variables in question are often atmospheric or oceanic flow fields whose state can be expressed in arbitrarily high dimension. While it is well understood that a low-dimensional bifurcation can be embedded in higher-dimensional dynamics, there has been little exploration of how the properties of that embedding affect the task of EWS detection. 

This paper aims to address the question of how EWS analysis of complex systems might be better informed by explicit attention to the embedding of a bifurcation in their state spaces. In particular, we investigate how inferences can be made in either direction between a hypothesis of tipping mechanism and the observation of critical phenomena. In cases where properties of a potential bifurcation can be hypothesized, we discuss how these hypotheses can be used to refine and robustify the EWS detection process (in the spatial aggregation step of the depiction in Box 1). Conversely, in the absence of prior knowledge about the tipping scenario, we demonstrate how EWS methodologies can be employed to make indirect observations of the support of the bifurcation embedding (bottom right panel of Box 1). Although previous EWS studies have in many cases made implicit use of assumed properties of the bifurcation in question (evidenced by their choice of measurement variables, sampling resolution, etc.), these hypotheses are rarely, if ever, contextualized in terms of their relationship to a bifurcation embedding. Likewise, the possibility of inferring embedding properties of a bifurcation directly from observed critical phenomena has gone largely unexplored in existing EWS literature.

We focus this study on local bifurcations taking place in high dimensional spaces, with particular attention to systems where spatiotemporally resolved measurements are made on one or more dynamical variables. The methods we introduce are entirely agnostic of the EWS detection algorithm employed. We present results obtained using a neural classifier model, as discussed in the Methods section, but the same analysis could equally be conducted using any other detection technique (albeit subject to the strengths and weaknesses of the chosen approach).

These cases are highly relevant to studies of climate tipping phenomena, where extensive geospatial data has been made available by the modern proliferation of remote sensing technology. Tipping points in the climate have been the subject of a great deal of recent interest amid a growing consensus that their potential abruptness and unpredictability places them among the gravest threats faced in a warming world \cite{Lenton2019,McKay2022,Wunderling2024,Steffen2018}. A central goal of this research is the identification of ``tipping elements"---components of the Earth system which may be prone to a runaway feedback mechanism when forced past some critical threshold \cite{Lenton2008}. Using the example of an abrupt shift observed in the annual West African monsoon system, we demonstrate how our method for inferring the structure of an embedded bifurcation can aid in this task. We employ EWS analysis to obtain a well-resolved spatial representation of the support of the critical dynamics directly from observation data, validating our results against those of a more conventional modeling approach.

\begin{tcolorbox}[skin=bicolor,
                    colback=blue!5!white,
                    colframe=white,
                    colbacklower=white,
                    left=0pt,
                    right=0pt,
                    top=8pt,
                    bottom=8pt,
                    width=\textwidth, 
                    enlarge left by=0mm,
                    boxsep=5pt,
                    arc=0pt,
                    outer arc=0pt,
                    float,
                    floatplacement=t,
                        ]
    \centering
    \hspace*{-1cm}
    \includegraphics[width=0.83\paperwidth]{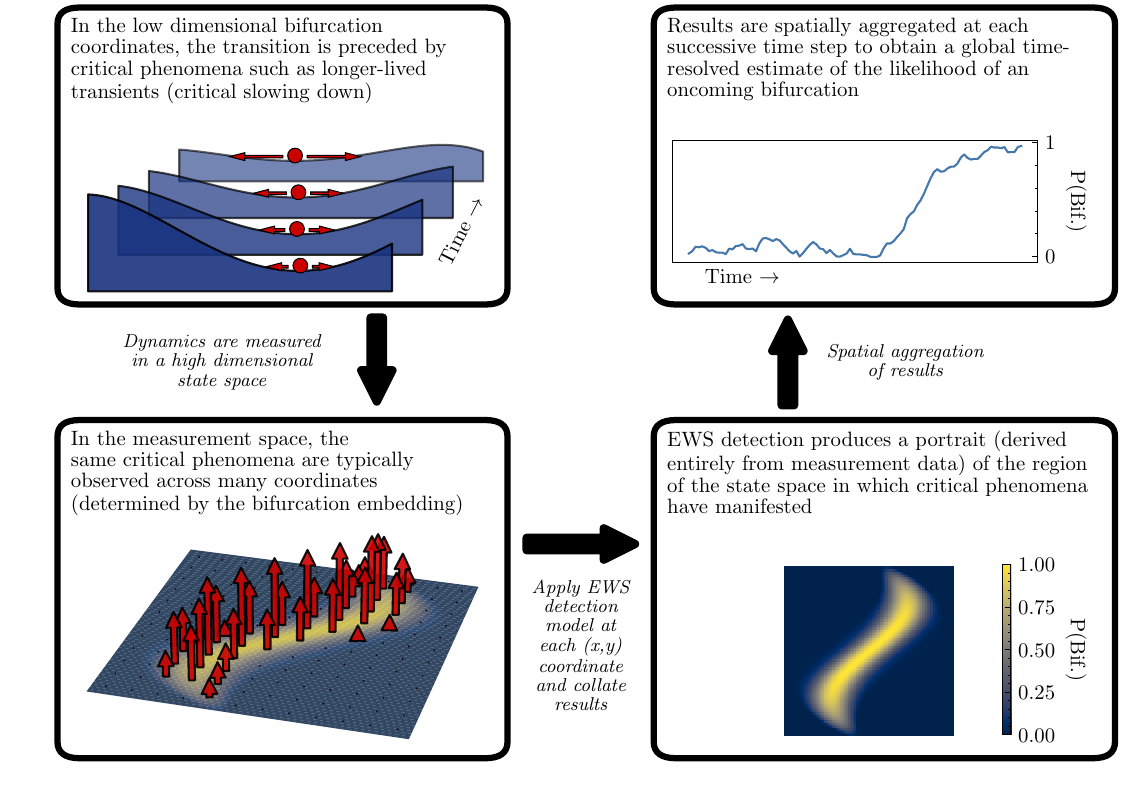}
    \tcblower
        \textbf{\textsf{Box 1:}} Conceptual schematic of embedded-bifurcation EWS analysis
\end{tcolorbox}
\subsection*{EWS of Embedded Bifurcations}

The basis of EWS detection techniques is the observation that many types of bifurcation are guaranteed to share certain characteristics, which can produce detectable signatures \cite{Dakos2008,Scheffer2009}. All local bifurcations in continuous time, for example, occur when an eigenvalue of the Jacobian matrix crosses the imaginary axis. A system in a stable equilibrium approaching a local bifurcation will find its basin of attraction growing increasingly shallow, resulting in reduced resilience and longer recovery times in response to small perturbations. This phenomenon of ``critical slowing down" (CSD) can be observed statistically in time series measurements, typically by examining trends in variance or autocorrelation \cite{Lenton2012}.

Bifurcations are taxonomized according to the different types of changes that the topological structure of a phase portrait can undergo. Any class of local bifurcation can be described by a normal form, a differential equation which minimally reproduces the bifurcation in as few dimensions as possible \cite{Strogatz2003}. All instances of a given type of local bifurcation are guaranteed to be topologically equivalent to the normal form near to the equilibrium. Crucially, this holds true even for high-dimensional systems: the bifurcation dynamics represented by the normal form are reproduced on a low-dimensional manifold embedded in the state space.


In practical terms, critical transitions in complex systems are often only approximated by local bifurcations. A system with dynamics on highly disparate time scales, for example, might admit an approximate representation for its slow manifold which reduces to the form of a bifurcation after averaging over the fast-scale dynamics. In these cases, the bifurcation embedding is a manifold on which observed dynamics are approximated imperfectly by the bifurcation normal form. It should also be noted that although CSD phenomena are produced by all types of local bifurcations, not all of these events are accompanied by the discontinuous, irreversible changes most commonly associated with the term "tipping point." Distinguishing between signatures of oncoming bifurcations which will or will not produce an abrupt equilibrium shift remains an open problem in EWS detection research. Some methods have shown promise, including neural time series classification models such as the one used in this study, but further investigation is needed \cite{Bury2020,Bury2021}. For the purposes of this paper we primarily concern ourselves with the detection of any bifurcation transition, continuous or otherwise. In practical terms, even those bifurcations which do not induce sudden and irreversible regime shifts may nonetheless lead to significant qualitative changes in a system's behavior on a relatively short time scale. While differentiating between the two remains an important goal, we emphasize that the results presented here do not offer any improvements on existing techniques to do so.

With respect to climate systems, the embedding (exact or approximate) of a bifurcation manifold within the state space is typically unknown. EWS detection methods are usually carried out using direct measurements of the system state under the assumption that the critical phenomena produced by the low-dimension bifurcation dynamics will be reproduced in the full state dynamics. This is, in many cases, a well-motivated assumption. Indications of an oncoming bifurcation are usually encoded in trends in variance, autocorrelation, skewness, and other statistical properties which are fairly robust to transformation: an increase in variance along the critical axis of the low-dimensional bifurcation system is very likely to correspond to an increase in variance of a measurement made on the high-dimensional state space (for any reasonably smooth and well-behaved embedding of the bifurcation). As a result, early warning for a tipping event can often be successfully measured even without any knowledge of the embedding. Nevertheless, any attempt to monitor EWS entirely agnostically of this relationship carries some risk: if one unknowingly measures the system state along a direction orthogonal to the bifurcation manifold, the signs of oncoming tipping can be completely obscured. This phenomenon of ``silent catastrophe" has been demonstrated in low-dimensional ecological models by Boerlijst et. al. (2013) \cite{Boerlijst2013}, but it leads to potentially greater difficulty in high dimensional cases, where large subspaces of unknown structure may be devoid of EWS information.

Applications of EWS detection methods to climate systems (and other systems of comparable complexity) can be carried out with varying levels of prior knowledge regarding the tipping mechanism in question. In some cases a well-supported reduced order model containing a critical threshold has already been developed, and a data-driven EWS approach is used to correct for any discrepancy between the simplified model's predicted tipping threshold and that of the fully realized system. Alternatively, EWS detection methods can be used without any prior expectation of the tipping dynamics in order to identify unforeseen mechanisms of criticality. Between these extremes lies a spectrum of possible levels of foreknowledge of the structure of the bifurcation manifold in question. Often the authors' conjectures about the embedding are not stated concretely, but are implicit in their choice of variables, sampling resolution, etc. In this paper we seek to foreground the importance of these assumptions to the execution and interpretation of EWS analysis.

\section*{Results}
\subsection*{Example: Hopf Bifurcation in 2D Fluid Flow}

\begin{figure}[!ht]
\centering
\includegraphics[width=\linewidth]{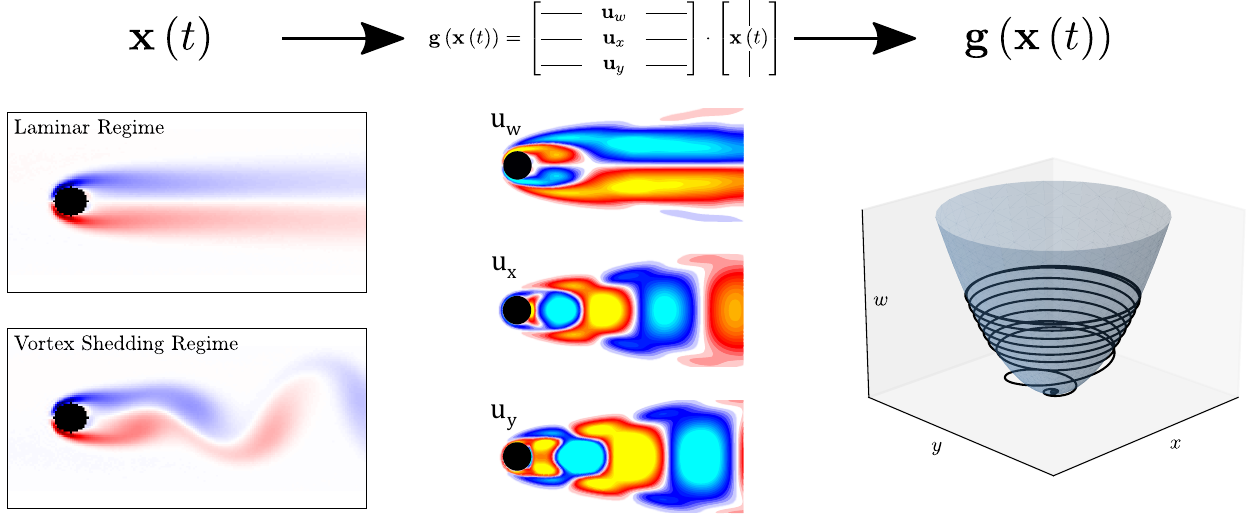}
\caption{Schematic of the embedding of the 2D Hopf bifurcation dynamics in the fluid system. The high-dimensional state $\mathbf{x}$ is (approximately) reduced to the bifurcation form by the embedding function $\mathbf{g}$. Left: vorticity plots of the full system state in its laminar equilibrium (top) and vortex shedding oscillation (bottom). Center: vorticity plots illustrating the spatial modes which form an approximate linear basis for the bifurcation manifold. Right: dynamics of the system projected into the 3D space spanned by these modes (black) overlaid on the 2D manifold on which the Hopf dynamics unfold (blue).}
\label{fig:vortex_schematic}
\end{figure}

As a demonstrative example, we present results for the application of EWS methods to a simulated fluid flow system undergoing a bifurcation. A fluid in two dimensions flows at uniform velocity towards a circular obstacle. By numerical simulation of the 2D Navier-Stokes equations, two distinct regimes can be observed: low flow rates produce a steady, laminar wake behind the obstacle, whereas high flow rates lead to a oscillatory solution known as a von Kármán vortex street, in which eddies of alternating vorticity are periodically shed past the obstacle. The abrupt transition of the system from a fixed equilibrium to a stable periodic orbit due to smooth variation of the flow rate bears all the hallmarks of a Hopf bifurcation, but the mechanism by which it occurs is not obvious: a Hopf bifurcation is produced by a cubic nonlinearity, whereas the Navier-Stokes equations governing the fluid system only contain terms up to second order. This apparent contradiction was resolved by the development of a mean-field model which isolates a slow timescale manifold on which dynamics approximate the cubic term of the Hopf normal form \cite{Noack2003,Brunton2016}. Details of this solution are presented in the supporting materials. 

This system presents a useful test case for EWS analysis. It undergoes an abrupt transition generated by a local bifurcation, but the embedding of the bifurcation dynamics in the observed state space is nontrivial. The mean-field model provides an analytic description of the embedding which we can use as a ground truth against which to validate our results. If we suppose that this analytic solution is not known, we consider two questions regarding the application of EWS methods:
\begin{itemize}
\item Can EWS detection applied to spatiotemporal measurements of the system offer any insight into the structure of the underlying bifurcation manifold?
\item How might the EWS methodology be improved and robustified by making hypotheses about properties of the bifurcation embedding (e.g. characteristic length scales)?
\end{itemize}

\noindent The following two sections present results to answer each of these questions.

\subsubsection*{Inferring structure of the bifurcation manifold}

\begin{figure*}[!ht]
\centering
\includegraphics[width=\linewidth]{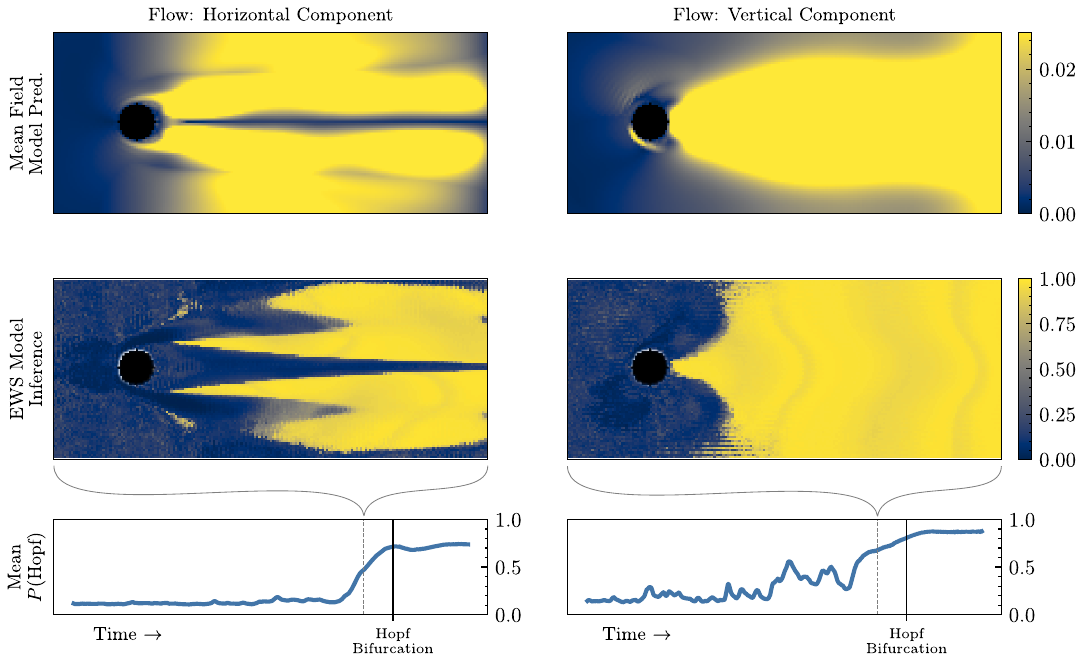}
\caption{Top: envelope of vortex shedding cycle in the 3D mean field model, projected onto the full state space of horizontal flow (left) and vertical flow (right). Middle: snapshot of the EWS detection model's spatially-resolved output probability of an oncoming Hopf bifurcation based on measurements of horizontal flow rate (left) and vertical flow rate (right). Bottom: spatially averaged model output over time as the system approaches its bifurcation (denoted by the black vertical line). The times of the middle snapshots are represented by the dashed vertical lines.}
\label{fig:vortex_shedding_pred}
\end{figure*}

In the absence of prior knowledge about the bifurcation embedding, EWS detection is generally carried out using data from direct measurements on the state space. Figure \ref{fig:vortex_shedding_pred} shows results for the neural EWS model applied to the simulated fluid flow system approaching its Hopf bifurcation. EWS probabilities are computed separately for the observed dynamics at each spatial coordinate on a sliding time window leading up to the critical point. It should be noted that this does forgo any information that might be encoded in spatial correlations, which has been shown to be useful for EWS detection in many spatiotemporal systems \cite{Kefi2014,Dylewsky2023}. We opt for this approach because treating each coordinate independently has the advantage of preserving spatial detail in the model output. Although we do not implement it here, a compromise between the two could be achieved by applying a spatiotemporal EWS model on a sliding spatial window over the data set. This would allow the model to make use of local spatial relationships, albeit at some cost to the spatial resolution of the results.

Results averaged over the spatial dimensions (bottom plot) show that the model has performed as expected: probability for the Hopf class remains low throughout most of the laminar flow regime and increases significantly as the system nears its vortex shedding transition. The bifurcation is successfully predicted using measurements of both horizontal and vertical components of the flow, with the latter slightly outperforming the former in its forecast horizon. This discrepancy may result from a difference in signal-to-noise ratio; dominant mean flow is in the horizontal direction, so the comparatively small perturbative excursions in which much of the EWS information is encoded may be easier to discern along the vertical axis.

Of greater interest is the spatial distribution of the early warning indication (Fig. \ref{fig:vortex_shedding_pred}, middle). For comparison, plots are generated to show the dynamics in the mean field model ($u_w$,$u_x$,$u_y$) lifted into the full state space (Fig. \ref{fig:vortex_shedding_pred}, top). At each location the maximum amplitude over the time series is plotted to form an envelope of the oscillations, which can be interpreted heuristically as a measure of how strongly dynamics on the bifurcation manifold are represented at that location. This offers a qualitative reference for the spatial structure we expect to ascertain from EWS analysis on measurements of the flow field.

On the whole, the predictions of the mean field model agree quite well with the EWS results. Positive classifications are observed in a tapered region encompassing the wake of vortices behind the cylinder. Moreover, the horizontal cleft directly behind the cylinder where the mean field model suggests no bifurcation dynamics should be observed is reproduced by the EWS classifier's output for measurements of the horizontal flow (Fig. \ref{fig:vortex_shedding_pred}, left).

The two approaches do not produce identical results, but this is not entirely unexpected. There are two likely reasons for the discrepancy:
\begin{enumerate}
\item The EWS results offer an improvement over the analytic approach
\item The EWS results are skewed by spatially heterogeneous perturbations to the bifurcation dynamics
\end{enumerate}

The first case is supported by the methods' differing assumptions: the analytic model used to generate the top plots is based on a linear mean field approximation to the full dynamics, whereas the EWS analysis does not rely on any simplifying assumptions. The spatial structure of the output from the latter may therefore represent an improvement over those simplified predictions. This highlights an important strength of the purely data driven EWS methodology, particularly when applied to complex systems which present a challenge to traditional modeling techniques.

The second reason for possible disagreement is based on the fact that the principal mechanism by which early warning signals manifest in dynamics is through the system's response to perturbations. Consequently, the ability to detect critical phenomena from measurement data depends not only on the system's endogenous dynamics as it nears bifurcation, but also on the magnitude and spectral properties of any forcing it is experiencing. In the case of this fluid flow system a very low level of white noise is applied at each step of the simulation, but perturbations to the bifurcation mechanics are primarily contributed by dynamics outside the subspace spanned by $(\mathbf{u}_w,\mathbf{u}_x,\mathbf{u}_y)$, which can be considered exogenous to the bifurcation. These dynamics, resulting from the discrepancies between the full Navier-Stokes evolution and its mean field approximation, have a spatial character of their own. This spatial heterogeneity is likely to cause the EWS model's sensitivity to vary by location in a manner that would not be captured by the mean field predictions plotted in Fig. \ref{fig:vortex_shedding_pred}. This phenomenon represents a potential impediment to inferring the structure of a bifurcation embedding purely through EWS analysis. It is worth noting, however, that the plots in Fig. \ref{fig:vortex_shedding_pred} are snapshots of model output well in advance of the onset of the bifurcation. Closer to criticality, the yellow portions of these plots expand considerably and fill nearly the entire region to the right of the cylinder. Even if sensitivity has been adversely impacted by nonuniform perturbative forcing, the model ultimately succeeds in correctly classifying its input as warning signals become more pronounced approaching the bifurcation. Furthermore, it is likely that a spatially convolutional EWS model could be trained on examples with heterogeneous forcing in order to mitigate this effect. As discussed previously, this would come at the expense of some spatial resolution in the output, but it is nonetheless an approach worth exploring in future work.

The recovery of this structure from EWS analysis offers considerable insight into the spatial character of the bifurcation embedding. While these results do not reproduce the full modal projection function of the mean-field solution, they illuminate a silhouette of the embedding as observed in the measurement space. This region is made up of those degrees of freedom along which critical phenomena preceding the bifurcation are detected (and, just as importantly, excludes those which do not exhibit EWS). While the result may seem fairly trivial for this example, where one could plausibly postulate the spatial form just from basic knowledge of the symmetries and physics of the system, the methodology readily extends to more complex cases. The model-free, data driven nature of the technique and the universality of the EWS signatures it identifies mean that it can be applied effectively even to systems in which dynamics unfold across a wide range of time and length scales and among many coupled variables.

\subsubsection*{Robustifying EWS detection using postulated properties of the bifurcation manifold}
We have demonstrated how EWS analysis can be employed to draw inferences about an unknown embedding of a bifurcation in some measurement space, but the same relationship can also be usefully applied in the other direction: if one can leverage prior knowledge to make certain assumptions about how a bifurcation will manifest in some system of interest, those can be used to robustify EWS detection to predict the critical transition. As an illustrative example, we again consider the Hopf bifurcation of the same 2D fluid system and assume that the mean field solution is not known. Even without knowledge of the ``proper" bifurcation basis, one can reasonably expect that the spatial embedding of the bifurcation will contain structures on the same order of magnitude as the characteristic length scale of the system, i.e., the diameter of the cylinder. This offers an additional constraint that can be placed on EWS analysis of the dynamics: if probability of an oncoming critical transition is computed separately at each spatial coordinate, one can evaluate the likelihood of an incipient vortex shedding bifurcation by searching for a \textit{coherent} rise in output probability across a region of the expected size.

\begin{figure*}[!ht]
\centering
\includegraphics[width=0.8\linewidth]{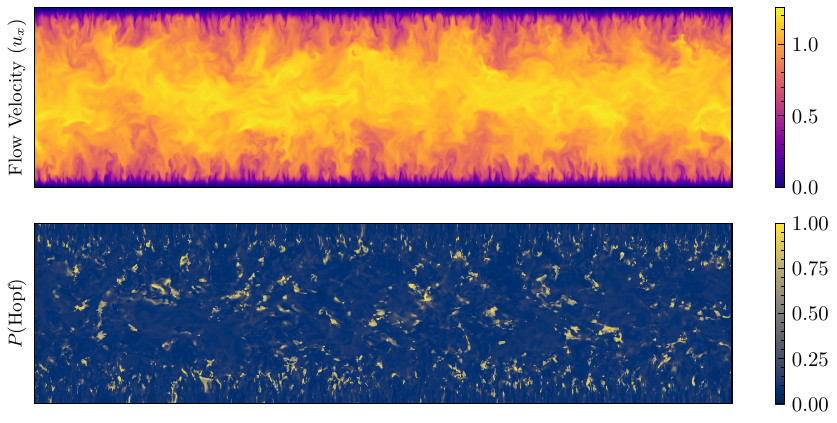}
\caption{Top: Snapshot of the $x$ component (i.e. along the horizontal) of the turbulent channel flow system (arbitrary units). Bottom: Snapshot of the spatially resolved output of the neural EWS model's Hopf class. The model's spatially-averaged output probability at any given time is quite low, but there are many localized regions in which it assigns a high likelihood of an oncoming bifurcation.}
\label{fig:JHTDB_Pred}
\end{figure*}

To illustrate the practical utility of this approach, we compare EWS detection results for the vortex shedding system to those for a simulation of turbulent fluid flow through a channel of finite width. Computations are carried out on a 2D slice of the 3D Turbulent Channel Flow data set made available by the Johns Hopkins Turbulence Database \cite{Perlman2007,Li2008,Graham2017}. Figure \ref{fig:JHTDB_Pred} (top) shows a snapshot of the channel flow velocity's $x$ component (along the direction of dominant flow). This example offers a useful control for EWS analysis; while the evolution of the system is (by construction) statistically stationary and thus not approaching any critical tipping point, turbulence produces rich and varied dynamics which present a significant challenge to the EWS classifier. Indeed, application of our neural EWS model to scalar time series measurements at each spatial location yields a positive classification in a nontrivial fraction of instances.

It would not be strictly correct to call these false positives, as local bifurcations may in fact be taking place: the fact that our model primarily assigns high probabilities to its Hopf class is particularly suggestive, given past work postulating that turbulence is generated by cascades of Hopf bifurcations \cite{Ruelle1971,Menon2016}. In practical terms, however, a negative classification is desired, as no tipping point (in the sense of large-scale and enduring qualitative regime shift of the dynamics) is ever observed. Spatially resolved output of the EWS model for this system, depicted in Fig. \ref{fig:JHTDB_Pred} (bottom), can be contrasted with those for the vortex shedding example (Fig. \ref{fig:vortex_shedding_pred}, middle row). In both cases the model at times forecasts an oncoming bifurcation with a high degree of certainty, but the degrees to which those affirmative classifications form spatially coherent structures on the scale established by system boundary conditions differ greatly.

\begin{figure*}[!ht]
\centering
\includegraphics[width=\linewidth]{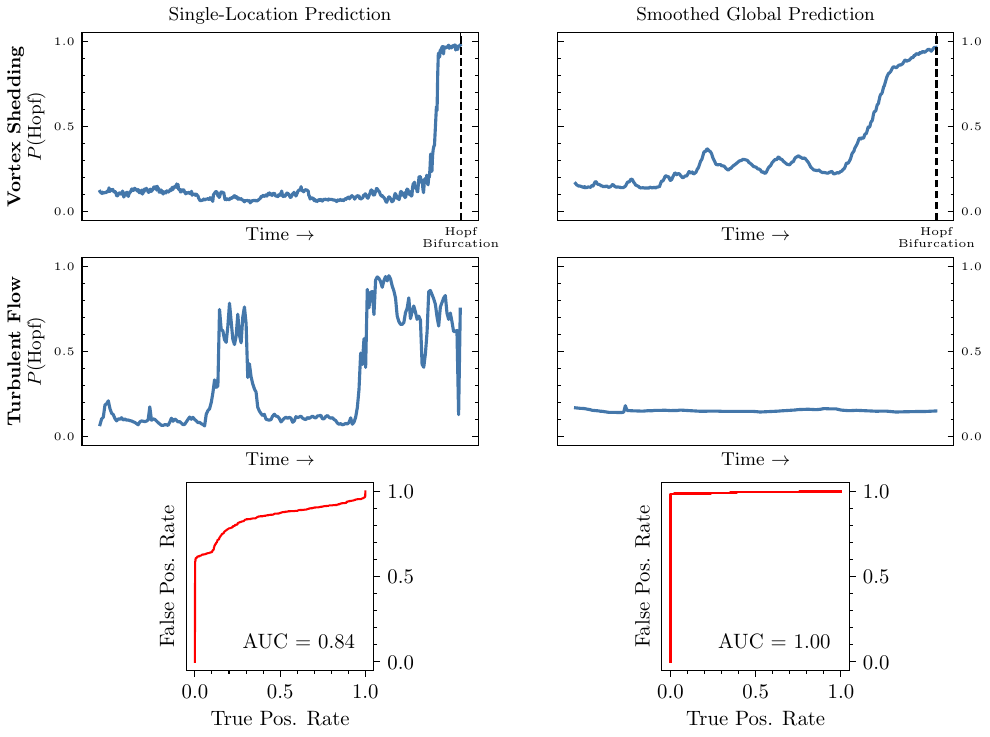}
\caption{Left: Time series of the neural EWS model's Hopf class output computed at a randomly chosen spatial location in the vortex shedding system (top) and turbulent flow (middle). Right: Time series predictions for the same systems computed by the spatially smoothed global method. For the vortex shedding system, the time of the true Hopf bifurcation is annotated by the dashed vertical lines. Bottom: ROC curves computed from an ensemble of randomly sampled spatial locations in both systems. In the smoothed case, this is carried out before computing the global maximum in order to allow for spatial sampling.}
\label{fig:JHTDB_Vortex_compare}
\end{figure*}

In Fig. \ref{fig:JHTDB_Vortex_compare} we present a simple example of how this difference might be operationalized for more robust EWS detection. Model output probability (for the Hopf class) over time for a single, arbitrary spatial location in each system is plotted on the left. Here we observe that the vortex shedding system yields a (true) positive classification shortly before the Hopf bifurcation, but the turbulent flow system also produces (effectively false) positive classifications at various times through its duration. To improve these results, we implement a simple postprocessing step: model outputs are smoothed along the spatial dimensions using a Gaussian convolution filter with width determined by the system's characteristic length scale (cylinder diameter and half channel width, respectively). This effectively obscures more localized excitations, allowing for an approximate global probability to be inferred from the maximum value of the resulting signal. These global probability metrics, plotted in Fig. \ref{fig:JHTDB_Vortex_compare} (right), produce the desired outcome: positive classifications for the turbulence case are entirely suppressed, while those for the vortex shedding case are preserved and even enhanced, arising further in advance of the bifurcation. This improvement is further demonstrated in the bottom row of Fig. \ref{fig:JHTDB_Vortex_compare}, where receiver operating characteristic (ROC) curves are produced from ensemble statistics over a large number of randomly sampled spatial locations in each system (in the right column, this sampling is carried out before the global maximum is computed). The area under the curve is observed to increase from 0.84 to 1.

This postprocessing methodology is expected to be quite adaptable. The example presented relies on an assertion of the expected spatial scale of the bifurcation, but a similar approach could be applied for an assumed characteristic time scale or for any other circumstance in which prior knowledge about the system can be used to constrain the search for an unknown embedding a high-dimensional measurement space. EWS detection methods have a documented proclivity for erroneous positive classifications \cite{Boettiger2012,Jager2019}, so any means of reducing such errors may constitute a considerable step forward for practical applications (although it should be noted that our method does not directly counter the sources of systematic error addressed in those references). 

\subsection*{Climate Application: West African Monsoon Jump}

To illustrate how these techniques can be applied to more complex systems, we present results for real-world climate data associated with the West African monsoon jump. This annual phenomenon, in which the latitude of maximum precipitation in the region of the Guinean coast shifts abruptly north in early July, has been analyzed extensively since it was first documented by Sultan and Janicot (2000) \cite{Sultan2000,Sultan2003,Hagos2007,Peyrille2016,Roehrig2013,Cook2019}. A system undergoing a rapid transition from one near-stationary state to another is highly suggestive of a critical tipping event. While a number of different mechanisms for the jump have been proposed, a recent work by Cook (2015) offers a compelling case in favor of this hypothesis \cite{Cook2015}. Specifically, Cook shows evidence that the phenomenon of atmospheric inertial instability plays a major role in undermining the stable conditions which precede the shift.

Inertial instability can be understood by examining perturbations to a purely geostrophic air flow, in which Coriolis and pressure gradient forces are balanced \cite{Holton2004}. If a parcel of air in this state is displaced along the horizontal direction orthogonal to the flow, the resulting imbalance between these forces generates a net acceleration proportional to the displacement vector. The constant of this proportionality acts as a stability criterion: if it is negative, displacements are met by a restoring force and the flow is stable, but if it is positive then any small perturbation will compound and the geostrophic equilibrium will be undermined. Cook numerically computes this criterion from reanalysis data and finds evidence that the stability threshold is crossed prior to the monsoon shift in a region associated with the Atlantic marine Intertropical Convergence Zone (ITCZ), $3^{\circ}$N to $6^{\circ}$N and  $25^{\circ}$W to $15^{\circ}$W, and subsequently also in a region along the Guinean coast ($3^{\circ}$N to $6^{\circ}$N and  $5^{\circ}$W to $5^{\circ}$E).

\subsubsection*{Inferring Spatial Structure of the Monsoon Jump}

\begin{figure*}[!ht]
\centering
\includegraphics[width=\linewidth]{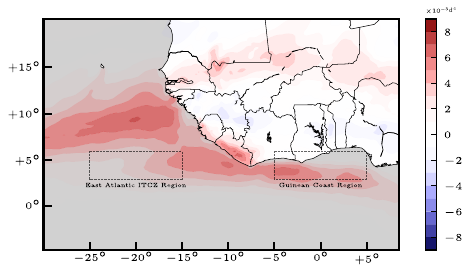}
\caption{Gradient of the neural model's fold class output (average daily rate of change) between June 15 and July 8 (mean over all years) for measurements of the meridional component of neutral wind at 10 m above the surface of the Earth. The detected onset of dynamic instability occurs most prominently just off the coast to the south and west of the mainland. The east Atlantic marine ITCZ and Guinean coast averaging regions used by Cook (2015) are labeled for reference.}
\label{fig:monsoon_pred_gradient}
\end{figure*}

We use this result as a point of comparison for the spatial embedding inferred by the EWS methodology described in the previous section. Climate data is obtained from the ERA5 global reanalysis for the years 1940 to 2022 \cite{Hersbach2020, Copernicus2023}. Hourly data is downsampled to daily means to remove day-night oscillations. Initial results make use of measurements of neutral wind at a height of 10m above the Earth's surface, with further discussion of variable selection to follow. Neutral wind is computed based on surface stress to reflect air flow in the idealized case of neutral atmospheric stratification. Following the reasoning of Cook, we focus our analysis on the meridional (north-south) component of wind: the dominant, approximately geostrophic flow in the (spatial and temporal) region of interest is primarily in the zonal (east-west) direction, so any critical phenomena produced by the proposed mechanism of inertial instability are expected to manifest in the system's response to meridional perturbations. As with the vortex shedding system, model predictions are computed separately for each spatial location on a sliding time window (of length 60 days). The observed dynamics of the ERA5 data are of course the product of the many overlapping mechanisms of diverse spatial and temporal scale which drive the evolution of the Earth's climate, any of which might produce signatures associated with critical transitions. In order to help isolate any signal connected with the dynamics connected to the monsoon shift phenomenon, we examine the gradient of model output averaged over a period of about 3 weeks prior to the expected onset time (roughly July 8, although it varies from year to year).

Results for this EWS analysis are plotted in Fig. \ref{fig:monsoon_pred_gradient}. As expected, the overall output probability of an oncoming critical transition rises significantly in the lead-up to the monsoon shift (i.e. the gradient is mostly positive). Moreover, the spatial features inferred by the EWS model agree reasonably well with the analysis of the Cook study, which demonstrates the onset of inertially unstable conditions in the averaging regions annotated in Fig. \ref{fig:monsoon_pred_gradient}. Cook does not provide any results more finely spatially resolved than this, but our model's success in producing a clear signal over the east Atlantic and off the Guinean coast (and virtually nowhere else) offers a fairly compelling qualitative validation of the methodology (as well as a corroboration of Cook's conclusions). This also highlights the utility of EWS-based inference of bifurcation embeddings in real-world systems: while the spatial structure discovered in the vortex shedding example was fairly obvious (essentially, the region corresponding to the cylinder's wake), the result here is quite nontrivial. It is both specific, because it points to a relatively small and well-defined area even with complex climate dynamics unfolding throughout the measured domain, and rich, in that it has a distinctive spatial structure that is not simply determined by the constraints and symmetries of the underlying system.

Interestingly, Cook's direct numerical estimation suggests that the stability threshold is only crossed concurrently with the monsoon shift in the Atlantic ITCZ, whereas the onset of instability on the Guinean coast occurs about a month later in early August. While it is encouraging that our model is able to detect early warning signals well in advance of this critical point, it does slightly complicate the interpretation of its spatial output: the loss of inertial stability migrates eastward over time, but if our goal is to understand the dynamical origins of the precipitation system collapse in early July, any phenomena subsequent to this event should ideally be omitted. The time horizon on which early warning signals can be detected is influenced by many factors and is difficult to estimate, however, so critical transitions which occur in close temporal proximity to one another cannot be easily disambiguated.

It should be noted that these results are generated from the model's output on the fold bifurcation class, which is not strictly consistent with the proposed inertial instability model. The formulation used by Cook and others describes a system in which perturbations from equilibrium are met by a restoring force directly proportional to the magnitude of displacement. This property produces the dynamics of a simple harmonic oscillator $\ddot{x} = -\alpha x$, which admits stable periodic orbits for $\alpha>0$ and no stable solutions for $\alpha<0$. The transition that occurs if $\alpha$ is smoothly tuned from one regime to the other does not belong to any of the local bifurcation classes on which the neural model was trained; rather, it is a global bifurcation in which the period of oscillation grows to infinity as the system approaches criticality. However, this does not mean that the neural classifier cannot be effectively used. Universal EWS features such as critical slowing down are still predicted to manifest, and the specific features which differentiate classes of bifurcation may be of negligible importance here. The simplified dynamical model 
for inertial instability may offer a useful description of the stability threshold the system crosses, but it is not expected to provide an accurate picture of dynamics within the stable regime (parcels of air will not be observed engaging in the spontaneous meridional oscillations that the equation would suggest). For this ERA5 data, the EWS model consistently assigns by far the strongest output probability to the fold bifurcation class, so these are the results we present in this section. This may suggest the existence of an alternate formulation in which the monsoon shift is well described by projection onto a low-dimensional manifold subject to dynamics given by the normal form of the fold bifurcation, but although this is entirely plausible, it is purely speculative. For the purposes of this analysis we simply note that many EWS signatures are shared across classes of bifurcation, so the neural model can reasonably be expected to behave as an effective binary classifier in the absence of more subtle distinguishing features. Fig. \ref{fig:monsoon_pred_gradient} could be generated by summing across all output classes, but in this case it would not look appreciably different.

\subsubsection*{Spatially-Robustified EWS for the Monsoon Jump}

\begin{figure*}[!ht]
\centering
\includegraphics[width=0.6\linewidth]{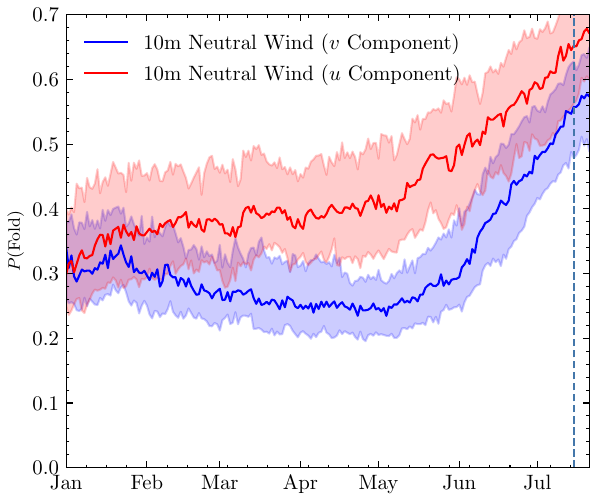}
\caption{Comparison between the neural model's EWS estimation for meridional ($v$) vs. zonal ($u$) flow. The thick lines represent median output over all years 1940-2022, and the shaded regions denote the 25th and 75th percentiles over the same set. Both signals increase significantly in the weeks leading up to the approximate date of the monsoon shift, which is denoted by the dashed vertical line.}
\label{fig:monsoon_var_comp_v10n_u10n}
\end{figure*}

As demonstrated in the vortex shedding example, an \textit{a priori} hypothesis of the characteristic length scale of a bifurcation embedding can be used to forecast an oncoming transition more accurately and reliably. In this case, we base this hypothesis on the inferred structure shown in Fig. \ref{fig:monsoon_pred_gradient}. The well-resolved spatial output offers a convincing portrait of the spatial extent of the tipping phenomenon, so we apply the Gaussian maximum method with a kernel width (FWHM) of approximately 130km: large enough to smooth out highly localized outliers, but small enough not to obscure the coherent signal which we assume to be associated with the monsoon jump. Results for this approach, plotted in Fig. \ref{fig:monsoon_var_comp_v10n_u10n} (blue), show a strong increase in the predicted probability of oncoming tipping as the monsoon jump approaches. Analogous results are also plotted for the same methodology applied to the zonal (east-west) component of 10m neutral wind (red). Estimated probabilities of oncoming transition are computed for each year in the range 1940-2022, and median results for the months leading up to the monsoon shift are plotted (computed over the region from $3^{\circ}$N to $12^{\circ}$N and  $30^{\circ}$W to $15^{\circ}$W). 

The strength of these results for both the $u$ and $v$ components of neutral wind merits a brief discussion regarding variable selection in these spatially-informed EWS methods. The results presented in this paper have focused primarily on inferring a bifurcation embedding across the degrees of freedom of a single continuum variable, but the same approach can be applied across multiple dynamic variables being measured simultaneously. Interpretation of multivariate results introduces some additional subtleties, however. We have thus far assumed that positive identification of early warning indicators along a given axis implies its non-orthogonality to the embedded bifurcation manifold, but this is not always the case. Consider a collection of system variables $\mathbf{x}$ with some bifurcation dynamics given by $\dot{\mathbf{x}} = \mathbf{f}\left(\mathbf{x}\right)$, and an ancillary variable space $\mathbf{y}$ evolving as $\dot{\mathbf{y}} = \mathbf{h}\left(\mathbf{x},\mathbf{y}\right)$. The dependence of $\mathbf{y}$ on $\mathbf{x}$ means that critical slowing down phenomena (e.g.) in $\mathbf{x}$ as it approaches its critical transition may propagate to $\mathbf{y}$, even though it is entirely exogeneous to the bifurcation system. The problem posed by these ``downstream" variables is typically negligible when comparing degrees of freedom of one continuum variable, since coupling is generally symmetric (i.e. the extent of influence of dynamics at location A on those at location B is expected to be equal to that of B on A). This assumption of symmetry may not hold when comparing two different variables, however, which can complicate the interpretation of EWS results. The criteria for the presence of EWS along a given direction may be better understood using the observability framework of control theory, as suggested in \cite{Negahbani2016}. For the sake of simplicity, we restrict our analysis in this work to variables that can reasonably be assumed to couple approximately symmetrically (e.g. the vertical and horizontal flow fields).

Even though the bifurcation model employed by Cook and others suggests that EWS should manifest in meridional dynamics, it is not surprising that the strongly coupled zonal flow field exhibits comparable predictive capability. Interestingly, examination of the spatial structure of the output for the zonal wind variable shows that its peak location occurs north of that of the meridional output by $5^{\circ}$--$10^{\circ}$ We hypothesize that this results from a progressive breakdown of the assumptions underpinning the inertial instability model for zonal geostrophic flow. In fact, there is an analogous stability condition that can be derived for zonal perturbations to a primarily meridional geostrophic flow \cite{Cook2015}, and it is likely that changes in ambient conditions across space and time lead to variability in the suitability of either approximation (which, in turn, determines the axis along which EWS signatures will be predominantly observed).

\begin{figure*}[!ht]
\centering
\includegraphics[width=0.6\linewidth]{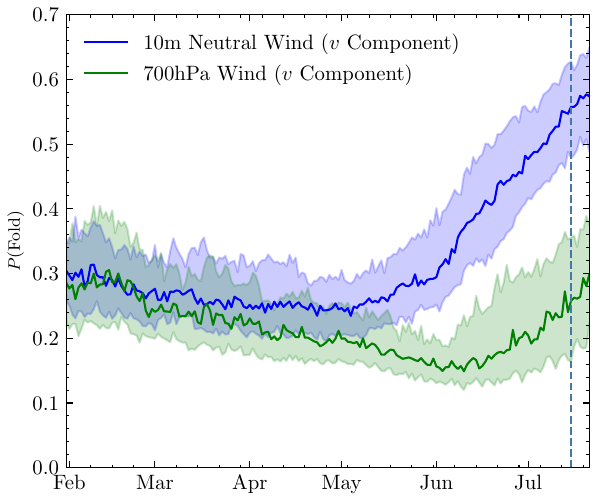}
\caption{Comparison between the neural model's EWS estimation for the two meridional wind variables. The thick lines represent median output over all years 1940-2022, and the shaded regions denote the 25th and 75th percentiles over the same set. The approximate date of the monsoon shift is denoted by the dashed vertical line. Although both signals exhibit a positive trend in the lead-up to the monsoon shift, the increase is much more prominent in the case of the 10m neutral wind.}
\label{fig:monsoon_var_comp_v10n_v}
\end{figure*}

A similar analysis motivates the choice of altitude of wind speed data used in this work. Our analysis of the West African monsoon system has made use of measurements of neutral wind measured at 10 m above ground level. This differs from the approach of the Cook study, which relies on wind measurements at the 700 hPa pressure level (approximately 3 km above sea level). Our choice of variable was prompted by the relative strength of EWS at these elevations; although both data sets exhibit a rise in the estimated probability of oncoming transition as the monsoon shift approaches, the 10 m data produces a much more pronounced signal. The observed difference in their respective model outputs is plotted in Fig. \ref{fig:monsoon_var_comp_v10n_v}.

We attribute this discrepancy to either of two likely factors. Firstly, as discussed previously, the task of EWS detection is easier when a system is observed responding to exogeneous perturbations. The 10 m data set is likely to exhibit much more prominent perturbations resulting from stresses associated with land or sea surface properties, so it may be the case that the same underlying critical dynamics are present at both elevations but are more readily observed at 10 m. Secondly, the disparity may reflect a legitimate spatial property of the inferred bifurcation embedding. Cook's choice to use 700 hPa data is motivated by the methodology for computing the criterion for inertial instability: this is the level at which observed air flow most closely fulfills the assumed conditions of geostrophic, zonally uniform flow. The formula used for direct numerical estimation of the stability criterion is only valid when these conditions are met. The EWS approach, however, makes no such demands. Our technique is purely data-driven, and thus not hindered by the possibility that the true mechanism of instability may not strictly adhere to the simplifying assumptions required for analytic analysis.

It is, of course, to be expected that the simple 1D model used to derive the stability criterion fails to capture the full complexity of the observed dynamics. Hagos and Cook (2007) suggest a number of potential complicating factors, including the complexity of continental background flow and the potential for non-geostrophic behavior \cite{Hagos2007}. It is the capacity of EWS analysis to elucidate the structure of an embedded tipping element in even the most complex and poorly-modeled systems that makes it a potent complement to more conventional modeling approaches. In a well studied system such the West African monsoon, the details of the spatial form represented in Fig. \ref{fig:monsoon_pred_gradient} can be cross-referenced with known climate phenomena in the region to make inferences about how the onset of instability actually manifests in real-world observation.

\section*{Conclusions}

EWS analysis can provide remarkably detailed insight into the embedding of bifurcation dynamics in arbitrarily high-dimensional systems. Its methodology is entirely data-driven and relies on detection of dynamical signatures expected to be universal across a very broad class of tipping behaviors, which makes it an extremely versatile and powerful technique. In this paper we have illustrated how EWS signatures can provide clues as to the variables and locations on which a bifurcation in a complex system takes place, and, conversely, how a well-motivated assumption about the properties of a bifurcation can be used to improve its advance detection. This bidirectional utility was demonstrated first on a relatively simple fluid system, for which a known analytic expression for the bifurcation embedding was available as a validation tool, and then on a real-world climate tipping phenomenon.

Although the methods presented are generically applicable to any high dimensional systems undergoing local bifurcations, their efficacy is constrained by the shortcomings of the chosen EWS detection technique---all EWS algorithms have different strengths and weaknesses with respect to accuracy, generalizability, and interpretability. Additionally, the issue of nonuniform detection sensitivity due to spatially heterogeneous perturbations may in some cases pose further difficulty to the data-driven inference process. In spite of these limitations, the demonstrated utility of these methods on both a simple synthetic system and an empirically observed climate phenomenon suggests that they have a practical role to play in the analysis of real-world bifurcation events.

The pragmatic potential of this approach is particularly salient in the study of Earth systems, where characterizing potential tipping points has become an increasingly urgent topic of research as the effects of climate change continue to intensify. The West African monsoon shift made for a suitable test case in this study because its mechanism is already somewhat well understood and has been observed repeatedly, allowing for validation of the results obtained. We hope, however, that the same methodology can prove useful in identifying as-yet unobserved tipping phenomena and in helping to refine predictive models of how these events might unfold, both in the climate and other complex nonlinear systems.

\section*{Methods}
\subsection*{EWS Detection Methods}
Results presented in this work are based on the application of an EWS detection model to time series data. A large variety of methodologies have been proposed for measuring signatures of oncoming critical transitions, diverse in both the types of transitions they are intended to forecast (local bifurcations, nonlocal bifurcations, phase transitions, etc.) \cite{Scheffer2009,Williamson2015,Hagstrom2021} and in the analytical means by which they do so (temporal or spatial computation of autocorrelation, moments of distribution, spectral methods, etc.) \cite{Lenton2012,Dakos2023}. Additionally, recent work has shown the potential of machine learning models to improve on these traditional statistical approaches by improving on accuracy and offering greater insight into the nature of the oncoming transition \cite{Bury2021,Deb2022,Dylewsky2023}. In this paper we employ a neural model of our own construction, based primarily on the design of Bury et. al. (2021) \cite{Bury2021}. The classifier takes as input a univariate time series of arbitrary length, and outputs a vector of probabilities assigned to each of the target classes: null (no upcoming bifurcation) or any of four types of local bifurcation (fold, Hopf, transcritical, and pitchfork).

We chose a deep learning approach for its classification accuracy and flexibility, but these methods are not without drawbacks. Neural networks are well known to be fairly opaque in their execution, sacrificing interpretability for performance. Furthermore, they tend to overstate their confidence in results for classification tasks, i.e. outputting class probabilities very close to 0\% or 100\% even when their demonstrated accuracy on validation tasks is considerably lower (though this can be mitigated somewhat through calibration methods such as temperature scaling) \cite{Guo2017}. We emphasize that none of the analysis in this paper is specific to any one EWS detection method. All results we present are produced by our neural classifier model, but the same methodologies could be equally effectively carried out using any other approach (modulo any potential shortcomings of the chosen technique).

\subsection*{Neural EWS Model}

\begin{figure*}[!ht]
\centering
\includegraphics[width=1\linewidth]{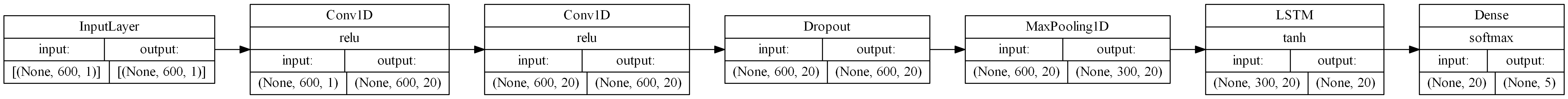}
\caption{Schematic representation of the neural network used for EWS detection. The dimensions labeled "None" refer to the user-specified batch size.}
\label{fig:model_schematic}
\end{figure*}

The neural EWS detection model used in this work is based on that of Bury et. al. (2021) \cite{Bury2021}, code for which is available at \url{https://github.com/ThomasMBury/deep-early-warnings-pnas}. Taking a one-dimensional time series as input, the network consists of two convolutional layers of 20 units each (kernel width 8) followed by an LSTM module with 20 memory cells. Between the CNN and LSTM layers there is a 10\% dropout layer and a 1D max pooling layer of width 2. Finally, a dense softmax layer is used to produce probability-like outputs for each of the target classes. A diagram of the model's architecture is shown in Fig. \ref{fig:model_schematic}.

Time series training data is generated by numerical integration of randomized equations of motion in 3 dimensions ($\dot{\mathbf{x}} = \mathbf{f}\left(\mathbf{x}\right)$, $x \in \mathbb{R}^3$). $\mathbf{f}$ is populated by sparsely randomized polynomial terms chosen from all possible interactions up to 5th order. The lower-order terms for each model are constrained such that it is guaranteed to reproduce a bifurcation normal form in the vicinity of the bifurcation, and randomized higher-order terms will not affect the topology of dynamics in this neighborhood. The dimension(s) normal to the bifurcation manifold are constrained to attract to it. Each instantiation of such a model is assigned one of four possible (local, codimension-one) bifurcation classes: fold (saddle-node), Hopf, pitchfork, or transcritical.

These equations are numerically integrated to simulate a bifurcation. At each time step, the deterministic polynomial dynamics are perturbed by a random additive noise term. This noise has standard deviation $\sigma$ its power spectral density per unit bandwidth is proportional to $f^{-\beta}$, where $\sigma$ and $\beta$ are sampled from log-normal and normal distributions, respectively, and held fixed for the duration of each run. The bifurcation parameter is assigned a randomized starting value, where it is held constant while the system is allowed to settle into a stable equilibrium. Once this is reached, the resulting equilibrium is used as an initial condition as integration proceeds with the bifurcation parameter varied linearly through its critical value. Null runs, in which no bifurcation takes place, are generated by varying the bifurcation parameter through some domain far from the critical threshold. Any runs which are observed to lose stability and blow up at any point prior to the intended bifurcation are discarded.

Runs produced in this fashion are then preprocessed. A scalar time series is extracted by measuring the 3D dynamics along a randomly chosen rotational axis. Bifurcation runs are then truncated from the right to discard all steps after the critical transition takes place, after which all runs are then randomly cropped from the left such that they vary in duration from $30$ to $600$ time steps (they are eventually padded from the left with zeros to obtain a uniform duration of $600$). A low level of additive white measurement noise is applied to each time series. Finally, each run is normalized to mean zero and unit standard deviation. Different additional detrending steps have been employed in other EWS detection approaches (e.g. subtraction of a Gaussian- or Lowess-smoothed trend), but we found that skipping this step did not adversely affect the accuracy or generalizability of trained models in the tests we ran.

Preprocessed data is then randomly sorted into train, test, and validation classes (85\%, 10\%, and 5\%, respectively). Each of the bifurcation classes are uniformly represented in the training population, and the null class makes up 25\% of the set. This latter proportion can be tuned to determine the model's overall likelihood of producing a positive classification; we chose a value that resulted in a conservative but not overly cautious bifurcation classification rate on the test systems of interest. Finally, the neural network is trained on batches of $32$ runs at a time using the Adam optimization algorithm and a categorical cross-entropy loss function. 

\bibliography{sample}

\section*{Author contributions statement}

Study conceived by DD, MA, and CTB. Computational results by DD. Manuscript and figures prepared by DD. All authors reviewed and edited the manuscript.

\section*{Data Availability}

Data for the turbulent channel flow system is made available by the Johns Hopkins Turbulence Database at \url{http://turbulence.pha.jhu.edu}. All data used for the West African Monsoon Shift analysis comes from the ERA5 global reanalysis \cite{Hersbach2020} provided by the Copernicus Climate Change Service (C3S) Climate Data Store (CDS) at \url{https://doi.org/10.24381/cds.143582cf} \cite{Copernicus2023}. Neither the European Commission nor ECMWF is responsible for any use that may be made of the Copernicus information or data it contains.

\section*{Funding}
The research was supported by NSERC Discovery Grants to MA (5006032-2016) and CTB (5013291-2019), and a DARPA Artificial Intelligence Exploration Opportunity Grant to MA and CTB (PA-21-04-02-ACTM-FP-012).

\section*{Competing interests}
The authors have no competing interests to declare.

\end{document}